# Design and simulation of a 12-bit, 40 MSPS asynchronous SAR ADC for the readout of PMT signal

Jian-Feng Liu (刘建峰)[1,2] Lei Zhao (赵雷)[1,2; 1)] Jia-Jun Qin (秦家军)[1,2] Yun-Fan Yang (杨云帆)[1,2]
Li Yu (于莉)[1,2] Yu Liang (梁宇)[1,2] Shu-Bin Liu (刘树彬)[1,2] Qi An (安琪)[1,2]

1 State Key Laboratory of Particle Detection and Electronics, University of Science and Technology of China, Hefei, 230026, China,

2 Department of Modern Physics, University of Science and Technology of China, Hefei, 230026, China

**Abstract:** High precision and large dynamic range measurement are required in the readout systems for the Water Cherenkov Detector Array (WCDA) in Large High Altitude Air Shower Observatory (LHAASO). This paper presents a prototype of 12-bit 40 MSPS Analog-to-Digital Converter (ADC) Application Specific Integrated Circuit (ASIC) designed for the readout of LHAASO WCDA. Combining this ADC and the front-end ASIC finished in our previous work, high precision charge measurement can be achieved based on the digital peak detection method. This ADC is implemented based on power-efficient Successive Approximation Register (SAR) architecture, which incorporates the key parts such as Capacitive Digital-to-Analog Converter (CDAC), dynamic comparator and asynchronous SAR control logic. The simulation results indicate that the Effective Number Of Bits (ENOB) with a sampling rate of 40 MSPS is better than 10 bits in an input frequency range below 20 MHz, while its core power consumption is 6.6 mW per channel. The above results are good enough for the readout requirements of the WCDA.

## 1 Introduction

The Large High Altitude Air Shower Observatory (LHAASO) project has been proposed for high energy gamma ray and cosmic ray detection [1-3]. One of the major components is the Water Cherenkov Detector Array (WCDA), which aims to survey the northern sky for the very high energy (VHE) gamma ray sources. The WCDA is a ground-based Extensive Air Shower (EAS) detector, which adopts 3600 Photomultiplier Tubes (PMTs) to collect the Cherenkov light produced by the secondary particles of EAS. To achieve wide energy spectrum observation with high energy resolution, the readout electronics for the WCDA are required to cover PMT signals in a range from single PhotoElectron (P.E.) to 4000 P.E. with precise time and charge measurement. A charge resolution of 30% @ 1 P.E. and 3% @ 4000 P.E. is required in the full dynamic range and the single-counting rate of a PMT can reach as high as 33-45 kHz [2].

For charge measurement, there are several methods applied in the readout electronics of high energy physics experiments, such as analog peak detection adopted in [4-7], digital peak detection adopted in [8-11], Time Over Threshold (TOT) principle adopted in [12-15], and waveform digitization based on Switched Capacitor Array (SCA) adopted in [16-19]. The analog peak detection has a relatively large dead time and the resolution of SCA-based waveform digitization is limited with high input signal frequency. In the readout electronics of the WCDA, we investigated the ASIC-based implementation of both TOT method and digital peak detection. The former can simplify the structure of readout electronics extremely while the latter has the advantages of better linearity and resolution. In this paper, we focus on the implementation of the latter. The charge measurement circuits are designed based on analog signal shaping, analog-to-digital conversion and digital peak detection logic.

---

*Supported by Knowledge Innovation Program of the Chinese Academy of Sciences (KJCX2-YW-N27), the CAS Center for Excellence in Particle Physics (CCEPP)

1) Email: zlei@ustc.edu.cn

In our previous work, a front-end ASIC [20] with the function of analog signal shaping has been designed. In this paper, we present the design and simulation of a prototype 12-bit 40 MSPS Successive Approximation Register (SAR) ADC to digitize the output signal of the front-end ASIC. The Effective Number Of Bits (ENOB) of the ADC aims to be better than 10 bits, which corresponds to an equivalent input noise of 0.69 mV for a 2 $V_{PP}$ full scale input range. In this condition, the charge resolution contributed by ADC for 1 P.E. signal is better than 5%, which is good enough for overall resolution requirement (30% @1 P.E.). The large number of readout channels demands low power consumption to release the pressure of cooling system on high altitude mountains. Therefore, the SAR architecture [21-22] is adopted for its simplicity and power-efficient in deep-sub-micron CMOS technology compared with pipelined architecture [8, 23]. This paper is organized as follows. The second section describes the overall ADC architecture. The third section presents the design and implementation of key circuit blocks. Simulation results are presented in the fourth section. Finally, the conclusions are given.

## 2 ADC architecture

The overall architecture of the actually implemented 12-bit ADC is depicted in Fig. 1. The ADC mainly consists of a Capacitive Digital to Analog Converter (CDAC), a dynamic comparator, and an asynchronous SAR control logic. A fully differential solution is used to minimize the common mode interference caused by charge injection, digital cross-talk and other disturbances and suppress the even harmonics. The CDAC is in responsible of sampling and charge redistributing. For sampling, the bootstrapped switches are employed while CMOS switches are used for charge redistributing. A high-speed and low-noise comparator performs the digitization of the CDAC's voltage differences. Both CDAC and dynamic comparator consume no static power. To further lower power consumption of clock tree and increase conversion speed, asynchronous control logic is implemented based on the dynamic latch structure. Besides, the ADC integrates bandgap bias, voltage reference buffer, and clock generator. The implementation of main ADC blocks is described in the following sections.

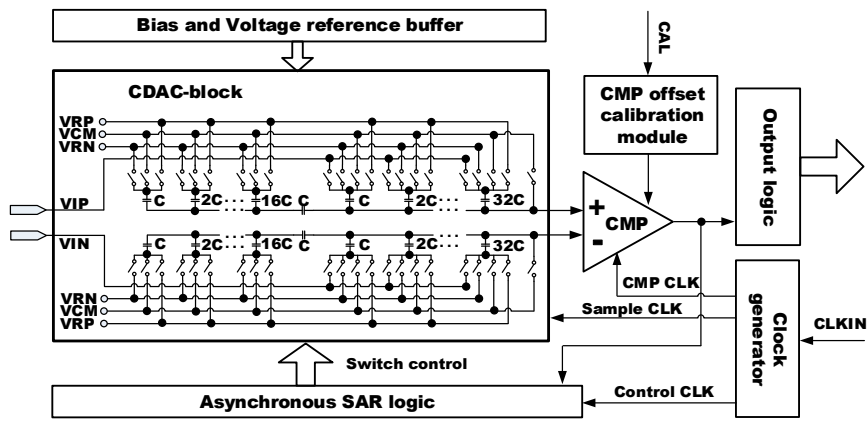

Fig. 1 Architecture of the PMT readout electronics in the WCDA of LHAASO.

## 3. Circuits implementation

### 3.1 Design consideration

The ADC performance is restricted by the trade-off between accuracy, power and speed. To achieve a 12-bit resolution, the noise from comparator and CDAC should be less than 0.5 mV (1 LSB). However, suppressing the noise of comparator and CDAC excessively will cause higher power consumption. To achieve 40 MSPS, we allocate about 20% of period time (5 ns) for sampling while the rest (20 ns) for conversion. This time allocation aims to minimize the total power consumption of sampling and conversion.

### 3.2 Bootstrapped switch

Bootstrapped switch [24-25] as shown in Fig. 2 is employed to reduce the variation of on-resistance and improve the switch linearity. In the sampling phase, the bootstrapped switch and CDAC perform the Sample/Hold (S/H) function. The differential solution and bottom sampling are adopted to minimize non-idealities (the clock feedthrough and charge injection), which can improve the dynamic performance of ADC. The on-resistance of the bootstrapped switch is optimized to be about 50 Ω when the tracking time is set to 5 ns. As shown in Fig.3 the ENOB of the S/H module is better than 12.5 bits with 1.8 pF input capacitance.

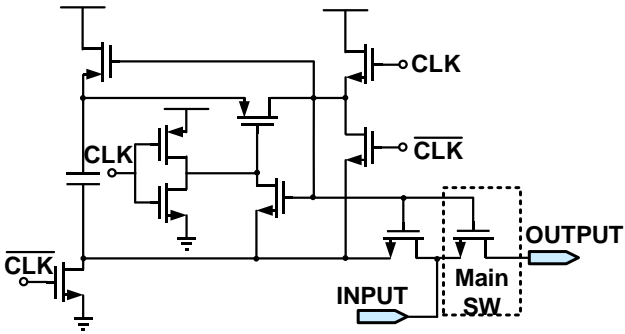

Fig. 2 Schematic of the bootstrapped switch

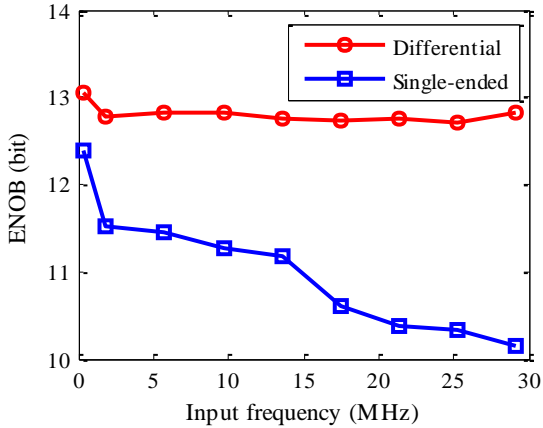

Fig. 3 ENOB simulation of results of S/H module.

### 3.3 Capacitive DAC

To further lower the power consumption, the CDAC incorporates split capacitor structure and $V_{CM}$-based switching scheme [26-27]. Compared with the conventional switching scheme, the $V_{CM}$-based method saves about 87% power. Besides, the 12-bit resolution can be achieved by employing 11-bit CDAC, which decreases the mismatch requirement and saves chip area. As shown in Fig. 1, in each CDAC array there are M=6 capacitors on the Most Significant Bits (MSB) side and L=5 capacitors on the Least Significant Bits (LSB) side. Metal-Insulator-Metal (MIM) capacitors are used to implement CDAC due to their good matching performance. As shown in Fig. 4, the Monte Carlo simulation of the CDAC indicates the DNL and INL are less than 1 LSB when the unit capacitance is about 28 fF. The total capacitance of CDAC MSB side is about 1.8 pF, which makes the thermal noise much smaller than the quantization noise of 12-bit resolution.

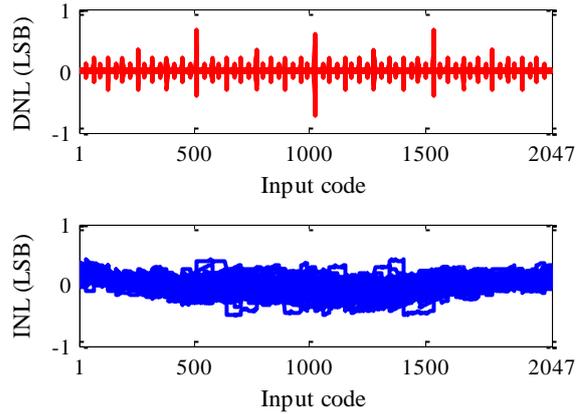

Fig. 4 Monte Carlo simulation of DAC in 50 runs.

Each bit of CDAC settling accuracy is determined by its time constant constituted by on-resistance of CMOS switch and capacitance of CDAC capacitor. The size ratio of CMOS switch in each CDAC bit is set according to the capacitance ratio to make each bit have the same time constant. The settling time of 0.5 LSB accuracy for 500 mV step pulse is optimized to less than 600 ps.

### 3.4 Dynamic comparator

The designed dynamic comparator [28-29] is presented in Fig. 5. To achieve better speed performance in 180 nm CMOS technology, structure without pre-amplifier is adopted. The operating speed of the comparator can be up to 1 GHz for 1 LSB imbalance.

Fig.6 shows the simulation results of the comparator noise obtained with Spectre transient noise simulation. Fitting the simulation results to a Gaussian cumulative distribution, we can obtain the RMS equivalent input noise is about 136 μV (<0.3 LSB).

Small areas of input transistor gates are chose to obtain a high speed. However, the smaller sizes of input transistors will cause larger mismatches, consequently, a large input referred offset. We use on-chip offset calibration [30] to compensate for this defect and allow a low power implementation of the comparator. Fig. 7 presents the Monte Carlo simulation results of offset calibration. Before calibration, the RMS value of offset is 2.7 mW while the RMS value is about 200 μV (<0.5 LSB) after calibration.

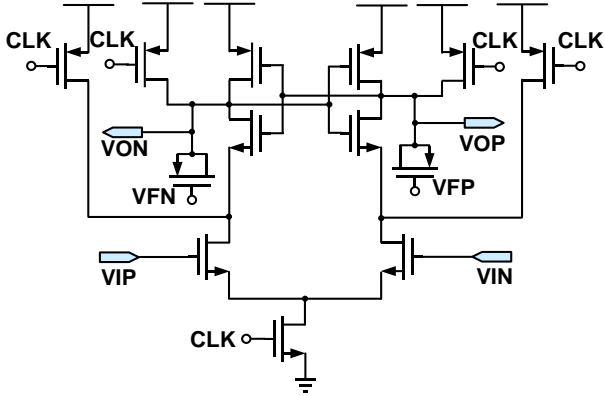

Fig. 5 Schematic of dynamic comparator.

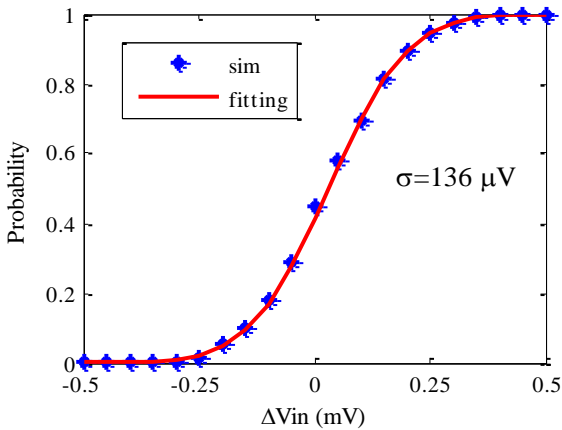

Fig. 6 Noise simulation of dynamic comparator.

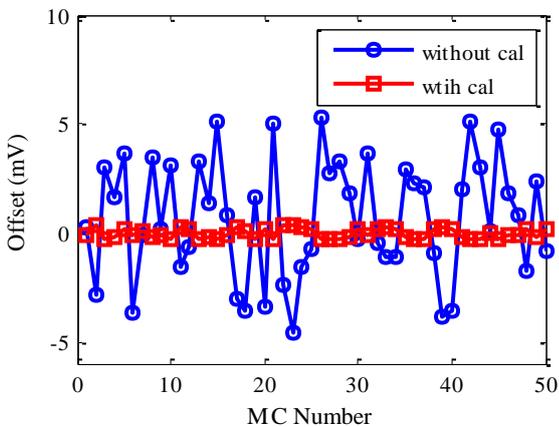

Fig. 7 Monte Carlo simulation of offset calibration.

### 3.5 Asynchronous SAR logic

Compared with synchronous SAR control logic, the asynchronous implementation [31] is much more efficient to achieve higher sampling rate and lower power consumption. For high-speed asynchronous logic, the conversion phase needs an internal high-frequency multi-phase clock to trigger the comparator. A gate-controlled ring oscillator (GCRO) [32], as shown in Fig. 8, is adopted to generate the multi-phase clock. The signal "CLKSD" shown in Fig.8 starts the conversion while the "$S_{12}$" stops the conversion. The voltage controlled variable delay is employed to adapt the comparator clock variation due to process corner.

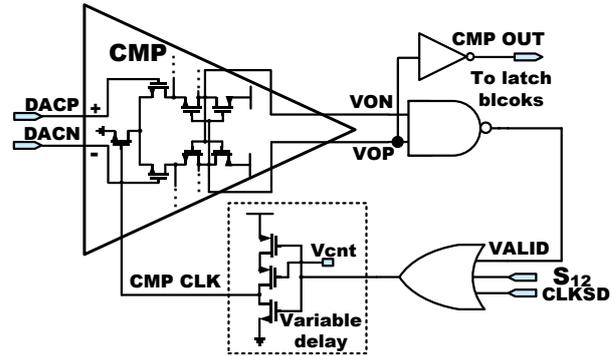

Fig. 8 Schematic of Multi-phase clock generator. "CLKSD" is a delay signal of sample clock and "$S_{12}$" is the flag signal pointing the end of conversion.

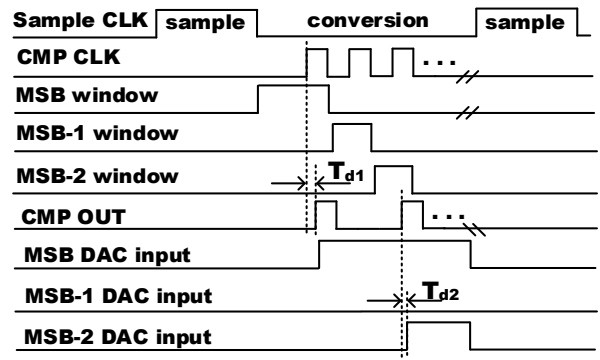

Fig. 9 Timing diagram of asynchronous SAR logic. "$T_{d1}$" is the delay of comparator decision and the "$T_{d2}$" is the latch delay of SAR logic

The conversion phase contains twelve periods of comparator clock, each of which includes DAC settling time, comparator decision time and SAR logic delay. Dynamic logic based on "window-opening" scheme [33], which can reduce the delay of the SAR logic, is incorporated to minimize the conversion time. The SAR logic adopts twelve latch blocks to generate CDAC digital input signal and 12-bit ADC output code. During bit cycling, each decision result of dynamic comparator is locked by one latch block when it is enabled by the corresponding "window" signal, which is generated by window generator. The timing diagram of SAR logic is presented in Fig. 9. In each period of sample clock ("Sample CLK"), the ADC operates in two phases: sample phase and conversion phase. The ratio of the two phases is adjusted by a Delay-Locked Loop (DLL). In the conversion phase, the sampled voltage is quantized under the control of SAR logic. When the "window" signal is valid, the corresponding latch block locks the decision result ("CMP OUT") of comparator and then the SAR logic passes it to CDAC to change its output value. The above cycle is repeated twelve times in each conversion phase and then the SAR logic resets the latch blocks for next conversion. Each bit delay ("$T_{d2}$" in Fig. 9) of the SAR logic is about 80 ps, which is much less than the conventional DFF-based SAR logic.

## 4. Simulation results

The prototype ADC was designed in Global Foundry 1P6M 180 nm CMOS technology. Fig. 10 shows the layout of the chip. It occupies a total area of 3×2 mm$^2$ and contains four ADC channels. A full-scale input range of 2 $V_{pp}$ is achieved when the power supply is 1.8 V and common mode voltage $V_{CM}$ is 1 V. Simulations were conducted to estimate the ADC performance and the results show as follows.

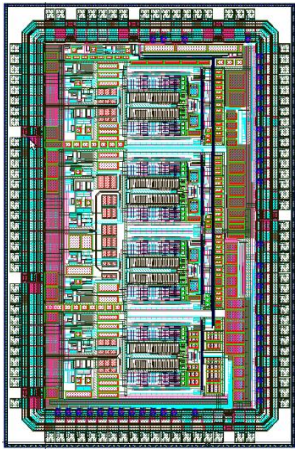

Fig. 10 Layout of the proposed ADC.

We input a differential sinusoidal signal to the ADC in the Spectre transient noise simulation environment. Fig. 11 plots the signal-to-noise-and-distortion ratio (SINAD) and ENOB values versus the input frequency while operating at 40 MSPS. The sine wave curve fit method [34] is adopted to reduce simulation time by using less samples than that of Fast Fourier Transform (FFT). A least-squared-error sine wave is fit to the output data of the ADC, and the resulting error of the two curves represents the power of noise and distortion. The signal power can be calculated by using the amplitude of the fit sinusoid.

Chip performance is summarized in Table 1. The ADC achieves 10.47-bit ENOB and the core part consumes 6.6 mW, which leads to a Figure-of-Merit (FOM) [35] of 116 fJ/conv-step. Chip verification will be done when the wafers come back.

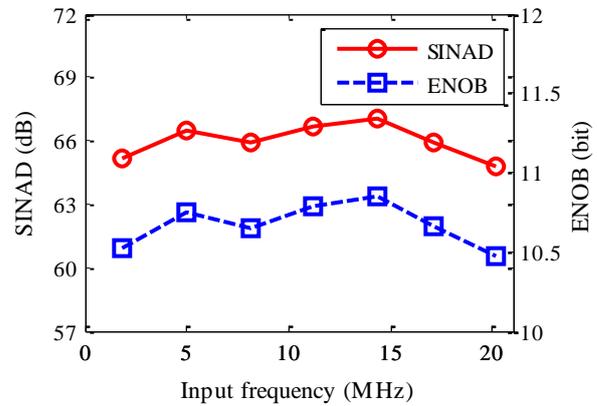

Fig. 11 Simulation of SINAD and ENOB versus input frequency.

Table 1. Performance summary

| Parameter | Value |
| --- | --- |
| Technology | 180 nm CMOS |
| Power Supply | 1.8 V |
| Sampling rate | 40 MSPS |
| Resolution | 12 bit |
| Full scale input | 2 $V_{pp}$ |
| ENOB | 10.47 bits @20.2 MHz |
| SINAD | 64.8 dB @20.2 MHz |
| Core power consumption | 6.6 mW |
| FOM | 116 fJ/conv-step |

We also simulated the ADC with the output signal of the front-end ASIC [20], which has a fast peaking time of 80 ns. As shown in Fig. 12, the samples of the ADC retain the information of the shaped signal well. The peak value of 1 P.E. shaped signal is 17.6 mV [20] and the equivalent input noise of the ADC is 0.5 mV, which means that the charge resolution contributed by this ADC is better than 5% for 1 P.E. input signal.

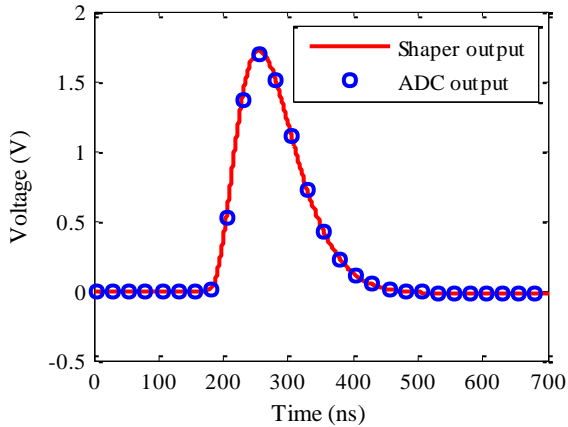

Fig. 12 Simulation with the shaped signal.

## 5. Conclusions

A 12-bit 40 MSPS asynchronous SAR ADC ASIC prototype in 180 nm CMOS technology is presented. Details regarding the key parts such as CDAC, dynamic comparator and asynchronous SAR logic are discussed. We have finished the design of the ADC which is now in fabrication, and conducted simulations to estimate its performance. The simulation results indicate that this ADC achieves 10.47-bit ENOB while its core power consumption per channel is 6.6 mW. The charge resolution contributed by this ADC is better than 5% @ 1 P.E., which is good enough for the readout requirement of LHAASO WCDA.